\documentclass[a4paper]{article}
\usepackage{color}
\usepackage[hyphens]{url}  % DO NOT CHANGE THIS
\usepackage{graphicx} % DO NOT CHANGE THIS
\usepackage{amsmath,bm}
\usepackage{url}
\usepackage{comment}
\usepackage{amsfonts}
\usepackage{todonotes}
\usepackage{hyperref}

\usepackage{textcomp}
\usepackage{fancyhdr}
\usepackage{authblk}
\usepackage{balance}
\usepackage{xcolor}
\usepackage{twocolceurws}
\usepackage{xcolor}
\usepackage{makecell}

\title{FBAdTracker: An Interactive Data Collection and Analysis Tool for Facebook Advertisements}

\author{
Ujun Jeong  \\ Arizona State University \\  ujeong1@asu.edu
\and

Kaize Ding \\ Arizona State University \\ kaize.ding@asu.edu
\and

Huan Liu \\ Arizona State University \\ huan.liu@asu.edu

}

\institution{}
\institution{}
\begin{document}
\maketitle

\section*{Abstract}
  The growing use of social media has led to drastic change in our decision making. Especially, Facebook offers marketing API which promotes business to target potential groups who are likely to consume their items. However, this service can be abused by malicious advertisers who attempt to deceive people by disinformation such as propaganda and divisive opinion. To counter this problem, we introduce a new application named FBAdTracker. The purpose of this application is to provide an integrated data collection and analysis system for current research on fact-checking related to Facebook advertisements. Our system is capable of monitoring up to date Facebook ads and analyzing ads retrieved from Facebook Ads Library\footnote{https://www.facebook.com/ads/library}.
  
\section*{Introduction}
Facebook advertising platform has been targeted by malicious users to spread disinformation such as propaganda and divisive opinion in the past years\cite{andreou2019measuring}. To cope with this situation, Facebook has regulated malicious users by their policies, but steps toward fighting these problems are not adequate because it is easy to reproduce advertisements on Facebook. This characteristic results in an enormous amount of ad content to traverse through the platform and makes it more hard to detect disinformation.

For the sake of advertisement transparency and mitigating disinformation, Facebook has released Facebook Ad library on May, 2018. With the API provided with this service, some studies has attempted to understand political advertisements on Facebook such as Ad Observatory\cite{edelson2019analysis, edelson2020security} and Ad Analyst\cite{silva2020facebook}. Despite this effort, there is still a limitation in that it is not easy for other researchers to scale up the range of ads to collect. Therefore, previous studies could look over some critical ads that should be considered since the API basically finds ads only for the queries requested.

To solve this limitation, we develop a new tool named FBAdTracker, which allows users to register their preferred advertisements to retrieve and summarize this comprehensive information in the view of advertisement and advertiser. The contribution of the FBAdTracker is two folds: (1) it provides a management tool for monitoring Facebook ads based on user's preference; and (2) it provides registered users with insightful statistics by conducting automatic analysis on the collected data. By providing this system, we aim to uncover diverse aspects of Facebook ads.

\section{FBAdTracker Description}
\subsection{System Architecture}
\begin{figure*}[t]
    \centering
    \includegraphics[scale=0.65]{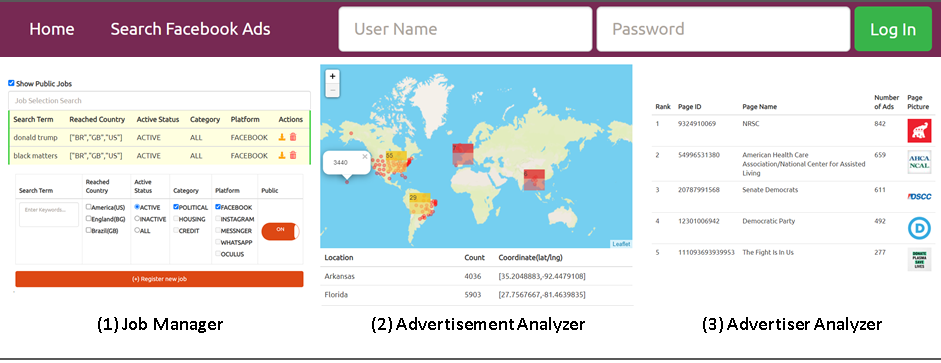}
    \caption{An illustrative example of FBAdTracker}
    \label{fig:one}
    \hfill
  \end{figure*}

The core of FBAdTracker is the database that monitors Facebook ads nearly in real-time. Due to the limited support of Facebook Ads Library API, it is impractical to retrieve all advertisements on Facebook. For this reason, our system solves this problem by registering multiple threaded requests of Facebook Ads Library API that persist until they are deleted. To support this idea, the system allows users to register their preferred advertisements to retrieve and FBAdTracker helps to interactively manage these multiple options such as search term, category, and etc. In the system, we call the combination of these options as a job for simplicity. Also, the system visualizes information available from Facebook ads so that researchers can use this comprehensive information for further analysis. As visually depicted in Figure \ref{fig:one}, the framework consists of three main components: (1) Job Manager (2) Advertisement Analyzer (3) Advertiser Analyzer to manage and analyze the database accumulated through interactions with users.

\textbf{Job manager} helps users to register their preferred advertisements to retrieve from Facebook Ads Library. When users enter this menu, they can select their customized options as suggested in Table\ref{tab1}. For example, a user can retrieve advertisements that contain the keyword `Trump' that sent to `Canada' with the category of `Social issues, elections or politics' only for the status of `Active' advertisements. Once these options are submitted, the system starts collecting data and also continuously update the recent advertisements not stored in the database. Especially, we can register a job as public and it allows other users to access data collected by this job. To overview these multiple jobs, users can list jobs available for them using the search bar and download data from the selected job in CSV format, and delete a job no longer needed to monitor.

\begin{table}[htbp]
\begin{center}
\def\arraystretch{1.35}
\scalebox{0.9}{
\begin{tabular}{|c|c|c|c|}
\hline
\textbf{Option}&\multicolumn{3}{|c|}{\textbf{Description}} \\
\hline
Search term&\multicolumn{3}{|c|}{Keywords to include for searching ads} \\
\hline
\makecell{Reached\\Country}&\multicolumn{3}{|c|}{\makecell{Countries where the ad has reached\\such as North America, England, and Brazil}} \\
\hline
\makecell{Active\\Status}&\multicolumn{3}{|c|}{\makecell{`ACTIVE' means eligible for delivery. \\`INACTIVE' means ineligible for delivery,\\and `ALL' includes both types}} \\
\hline

Category&\multicolumn{3}{|c|}{\makecell{Currently, `Political and Issue' can be \\ selected as a type of ads to search}} \\
\hline

Platform&\multicolumn{3}{|c|}{\makecell{Facebook, Instagram, Messenger, Whats-\\app, and Oculus are used for searching ads}} \\
\hline
\end{tabular}}
\caption{Required options to register a job}
\label{tab1}
\end{center}
\end{table}

\textbf{Advertisement Analyzer} visualizes the regional distribution of Facebook ads in the database. Users can select a period of time to analyze and select each spot to check the number of advertisements. Due to the lack of details on location given by Facebook Ads Library, the system helps to find the right geospatial information by looking up coordinates(latitude and longitude) and cluster the region by grouping the nearest coordinates on the map. Moreover, detailed information is shown below the map. Each location is ranked in ascending order to show top locations where advertisements have reached the most in the database.

\textbf{Advertiser Analyzer} ranks advertisers based on the number of advertisements. Users can select a period of time and the system collects every advertisement according to the selection. Then, the system groups them by Page ID which is a unique number assigned to the advertiser's page. On top of that, the system also collects the profile image on the advertiser's page using Graph API in Facebook. When the system gets URL for the profile image, it downloads binary encoded file. This information is ranked in ascending order to show top advertisers who have published the most advertisements in the database.

\subsection{Data Description and Policy}
As described in Table \ref{tab2}, there are diverse types of attributes that this system crawls using Facebook Ads Library API\footnote{https://www.facebook.com/ads/library/api}. Users can use this data within the purpose of their research according to the Facebook development policy\footnote{https://developers.facebook.com/terms/}. It is also allowed to share data obtained from Facebook Ad Library with other researchers or journalists if they have completed Facebook.com/ID confirmation and created a Facebook developer account. To this end, the review process will be done by the system manager when a user signs up. 

\begin{table*}[htbp]
\def\arraystretch{1.35}
\begin{center}
\scalebox{0.9}{
\begin{tabular}{|c|c|c|c|}
\hline
\textbf{Attribute}&\multicolumn{3}{|c|}{\textbf{Description}} \\
\hline
Id&\multicolumn{3}{|c|}{ID for the archived ad object} \\
\hline
Page Id&\multicolumn{3}{|c|}{ID of the Facebook Page that ran the ad} \\
\hline
Page Name&\multicolumn{3}{|c|}{Name of the Facebook Page which ran the ad} \\
\hline
Creation Time&\multicolumn{3}{|c|}{UTC time when someone created the ad. This is not the same time as when the ad ran} \\
\hline
Body&\multicolumn{3}{|c|}{The text which displays in the ad. Typically 90 characters} \\
\hline
Link Caption&\multicolumn{3}{|c|}{If an ad contains a link, the text that appears in the link} \\
\hline
Link Description&\multicolumn{3}{|c|}{If an ad contains a link, any text description that appears next to the link} \\
\hline
Link Title&\multicolumn{3}{|c|}{If an ad contains a link, any title provided} \\
\hline
Snapshot URL&\multicolumn{3}{|c|}{\makecell{URL link which displays the archived ad including images and videos from the ad}} \\
\hline
Spend \& Currency &\multicolumn{3}{|c|}{An amount of money spent running the ad as specified ISO currency code} \\
\hline
Funded Entity&\multicolumn{3}{|c|}{The name of the person, company, or entity that provided funding for the ad.} \\
\hline
Delivery Time&\multicolumn{3}{|c|}{UTC time when an advertiser wants Facebook to start \& end delivering an of the ad} \\
\hline
Impression&\multicolumn{3}{|c|}{A string containing the number of times the ad created an impression} \\
\hline
Potential Reach&\multicolumn{3}{|c|}{An estimate of the size of the audience that's eligible to see this ad} \\
\hline
\makecell{Regional \\distribution}&\multicolumn{3}{|c|}{\makecell{Regional distribution of people reached by the advertisement. \\Provided as a percentage and where regions are at a sub-country level}} \\
\hline
\makecell{Demographic \\distribution}&\multicolumn{3}{|c|}{\makecell{The demographic distribution of people reached by the advertisement.\\ Provided as age ranges and gender}} \\
\hline
\end{tabular}}
\caption{Description of attributes used in the system based on Facebook Ads Library API}
\label{tab2}
\end{center}
\end{table*}

\section{Conclusion and Future work}
Facebook advertisement is a new battlefield in terms of disinformation detection. There are often newly emerging issues that are hard to judge and malicious advertisers use this limitation to instigate the public. The proposed system contributes to this problem by providing the first integrated platform for data collection and analysis that can deal with the most recent advertisements on Facebook.

For the future work, it is demanded to develop a strong fact-checking system which can work with comprehensive information on Facebook Ads Library. To this end, we can utilize advertisements collected by our system to understand suspicious activities on Facebook Advertisement. The overall suspicious ads detection can be formulated in the bottom-up
approach where the lowest level is detecting individual ads that spread disinformation and the
top-level is detecting pages and advertising entities that are frequently generating disinformation. We hope our work can be meaningfully used for researches on fact-checking area.

\section{Acknowledgments}
This work is supported in part by ONR through grants N00014-21-1-4002.
\bibliographystyle{splncs04}
\bibliography{SBP2021}

\begin{thebibliography}{1}
\providecommand{\url}[1]{\texttt{#1}}
\providecommand{\urlprefix}{URL }
\providecommand{\doi}[1]{https://doi.org/#1}

\bibitem{andreou2019measuring}
Andreou, A., Silva, M., Benevenuto, F., Goga, O., Loiseau, P., Mislove, A.: Measuring the facebook advertising ecosystem. In: NDSS 2019-Proceedings of the Network and Distributed System Security Symposium. pp. 1--15 (2019)

\bibitem{edelson2020security}
Edelson, L., Lauinger, T., McCoy, D.: A security analysis of the facebook ad library. In: 2020 IEEE Symposium on Security and Privacy (SP). pp. 661--678. IEEE (2020)

\bibitem{edelson2019analysis}
Edelson, L., Sakhuja, S., Dey, R., McCoy, D.: An analysis of united states online political advertising transparency. arXiv preprint arXiv:1902.04385  (2019)

\bibitem{ghosh2019analyzing}
Ghosh, A., Venkatadri, G., Mislove, A.: Analyzing political advertisers’ use of facebook’s targeting features. In: IEEE Workshop on Technology and Consumer Protection (ConPro’19) (2019)

\bibitem{kumar2011tweettracker}
Kumar, S., Barbier, G., Abbasi, M., Liu, H.: Tweettracker: An analysis tool for humanitarian and disaster relief. In: Proceedings of the International AAAI Conference on Web and Social Media. vol.~5 (2011)

\bibitem{kumar2014twitter}
Kumar, S., Morstatter, F., Liu, H.: Twitter data analytics. Springer (2014)

\bibitem{morstatter2013understanding}
Morstatter, F., Kumar, S., Liu, H., Maciejewski, R.: Understanding twitter data with tweetxplorer. In: Proceedings of the 19th ACM SIGKDD international conference on Knowledge discovery and data mining. pp. 1482--1485 (2013)

\bibitem{silva2020facebook}
Silva, M., Santos~de Oliveira, L., Andreou, A., Vaz~de Melo, P.O., Goga, O., Benevenuto, F.: Facebook ads monitor: an independent auditing system for political ads on facebook. In: Proceedings of The Web Conference 2020. pp. 224--234 (2020)

\bibitem{zafarani2014social}
Zafarani, R., Abbasi, M.A., Liu, H.: Social media mining: an introduction. Cambridge University Press (2014)

\end{thebibliography}
\nocite{*}
\end{document}